\documentclass{ws-procs975x65}

\usepackage{amsfonts}

\usepackage{amscd}

\newcommand{\beq}{\begin{equation}}
\newcommand{\eeq}{\end{equation}}
\newcommand{\ben}{\begin{eqnarray}}
\newcommand{\een}{\end{eqnarray}}

\def\1{\'{\i}}

\def\>#1{{\bf #1}}

\def\ie{{\rm i}}

\begin{document}

\title{Conformal group with two observer independent scales.}

\author{Nicola~Rossano~BRUNO}

\address{Dipartimento di Fisica, Universit\`a di Roma ROMA TRE, and INFN Sez.\ Roma Tre,\\
Via Vasca Navale 84, 00146 Roma, Italy\\
Departamento de F\'isica,
Universidad de Burgos, Pza.\ Misael Ba\~nuelos s.n., \\
09001 Burgos, Spain }


\maketitle

\abstracts{
The Poincar\'e sector of a
recently deformed conformal algebra
is proposed to describe,
after the identification of the deformation parameter
with the Planck length, the symmetries of a new relativistic theory
with two observer-independent scales (or DSR theory). Also a new
non-commutative space-time is proposed. It is  found that momentum space
exhibits the same features of the DSR proposals preserving
Lorentz invariance in a deformed way. The space-time sector is a generalization
of the well known non-commutative $\kappa$-Minkowski space-time
which however does not preserve Lorentz invariance, not even in
the deformed sense. It is shown that this  behavior could be
expected in some attempts to construct DSR theories starting
from the Poincar\'e sector of a deformed symmetry larger
than Poincar\'e  symmetry, unless one takes
a variable Planck length. It is also shown that
the formalism can be useful in analyzing the
role of quantum deformations in the ``AdS-CFT correspondence".}

One of the most studied applications of quantum groups in physics is
in the description of deformed spacetime symmetries  that generalize    classical
Poincar\'e kinematics beyond  the Lie-algebra level.
Well-known examples
are the $\kappa$-Poincar\'e~\cite{LukierskiRuegg1992}
and the quantum null-plane (or light-cone)
Poincar\'e~\cite{nulla} algebras.
The deformation parameter has been interpreted
as a fundamental scale which may be related with the Planck length.
In fact, these results
can be seen as   different attempts to  develop new approaches
to physics at the Planck
scale~\cite{Majida}.
It has been also recently conjectured that the $\kappa$-Poincar\'e   algebra
may provide the basis for a so-called  doubly special relativity
(DSR)~\cite{Amelino-Camelia:2000,MagueijoSmolin,Amelino-Camelia:2002vy,Kowalski-Glikman:2002we}
in which the deformation parameter/Planck length is viewed as an
observer-independent length scale completely analogous to the familiar observer-independent
velocity scale $c$, in such a manner that (deformed) Lorentz invariance
is preserved~\cite{Bruno:primo,Bruno:2002wc,Lukierski:2002df}.
In this talk I want to review briefly some related results
which have been presented in greater detail
in~[\refcite{spainDSR,spainMink}]. These results provide an analysis
of the Poincar\'e sector of a manageable deformation of $so(4,2)$
introduced in~[\refcite{Herranz:2002fe}] together with its dual
in order to extract some physical implications
on the associated non-commutative Minkowskian spacetime.

If $\{J_i, P_\mu=(P_0,\> P), K_i,   D\}$ denote
the generators of rotations, time and space translations, boosts
and dilations, the non-vanishing deformed commutation rules of   $U_\tau({\mathcal WP})$  are given by~\cite{Herranz:2002fe}:
 \beq
\begin{array}{lll} [J_i,J_j]= \ie\, \varepsilon_{ijk}J_k  &\qquad
[J_i,K_j]=\ie\, \varepsilon_{ijk}K_k  &\qquad [J_i,P_j]=\ie\,
\varepsilon_{ijk}P_k   \\[2pt] [K_i,K_j]=-\ie\,
\varepsilon_{ijk}J_k
 &\qquad   [K_i,P_0]=\ie\, {\rm e}^{-\tau P_0} P_i  &\qquad
 [D,P_i]=\ie\,P_i  \\[2pt]
   \displaystyle{ [K_i,P_i]=  \ie\,
   \frac{{\rm e}^{\tau P_0}-1}{\tau} }  &\qquad
\displaystyle{[D,P_0]=\ie\, \frac{1-{\rm e}^{-\tau P_0}}{\tau}}  &
\end{array}
\label{ba}
\eeq
where hereafter  we   assume   $\hbar = c =1$,   sum over
repeated indices,  Latin indices $i,j,k=1,2,3$, while Greek
indices
$\mu,\nu=0,1,2,3$ and $\tau$ is a {\em real}
deformation parameter. By means of familiar techniques developed
for the study of DSR theories,  finite  boost
transformations were obtained in~[\refcite{spainDSR}] as well as a detailed
analysis of the range of  the boost parameter. It was found that the range of
the rapidity depends on the
sign of the deformation parameter and, in general, the behavior
of the rapidity, energy and momentum differs from DSR theories
based on $\kappa$-Poincar\'e.  For positive deformation parameter, the
situation is rather similar to the undeformed case, while for
negative deformation parameter, the range of the rapidity is restricted between
two values which are asymptotes for the energy. These
values determine a  maximum momentum that depends not only  on
the deformation parameter (as in the DSR
of [\refcite{Amelino-Camelia:2000,Bruno:primo}])
but also on the deformed mass of the particle. Furthermore two
proposals for position and velocity operators were presented: one
of them provides a variable speed of light for massless
particles, while the other one gives rise to a fixed speed of
light and also introduces a new type of generalized uncertainty
principle.

From a dual quantum group  perspective, when the quantum
spacetime coordinates $\hat x^\mu$ conjugate to
the  $\kappa$-Poincar\'e momentum-space $P_\mu$ (translations)
are considered, the  non-commutative $\kappa$-Minkowski spacetime
arises~\cite{Maslanka,Majid:1994cy,Zak1,Zak2,gacdual}. More general non-commutative
Minkowskian spacetimes can be described by means of the following Lie algebra
commutation rules~\cite{Lukierskid}:
\beq
[\hat x^\mu, \hat x^\nu]=\frac 1{\kappa} (a^\mu\hat x^\nu-a^\nu\hat x^\mu),
\label{aa}
\eeq
where $a^\mu$ is a {\em constant}
four-vector in the  Minkowskian space.
In [\refcite{spainMink}] the dual quantum group
of (\ref{ba}) was analyzed
making use of the quantum $\mathcal R$-matrix
(the well known FRT approach~\cite{Faddeev:ih})
and giving rise to the new non-commutative spacetime
\beq
[\hat x^\mu , \hat x^\nu ] =\tau
  \left(\hat \Lambda^\nu_{0}(\hat\xi) \hat x^\mu -
\hat \Lambda^\mu_{0}(\hat\xi) \hat x^\nu   \right) ,
\label{fa}
\eeq
which  can be seen as a   generalization of (\ref{aa})
through   $a^\mu\to \hat \Lambda^\mu_{0}(\hat\xi)$.
Since $\hat \Lambda^\mu_{0}$
(which are formal quantum Lorentz entries)
only depend on the quantum boost parameters
and  the quantum rotation coordinates $\hat\theta^i$
do not play any role in the spacetime non-commutativity,
the isotropy of the space is thus preserved.
The relations (\ref{fa})    show that  different observers in
relative motion with respect to quantum group transformations
have a different perception of the
spacetime non-commutativity, {\em i.e.},
the equivalence of all the inertial frames is lost.
Nevertheless, in this context, covariance under quantum group
transformations is ensured by construction.

We also stress that if the following new space variables $\hat X^i$    in the
$(3+1)$D spacetime (\ref{fa}) are considered
\beq
\hat x^0 \rightarrow  \hat x^0  ,
\qquad
 \hat x^i \rightarrow   \hat X^i=
\hat x^i\hat \Lambda^0_0-\hat x^0\hat\Lambda^i_0   ,
\eeq
the   transformed commutation rules for the quantum spacetime are given by
\beq
[\hat X^i,\hat x^0]=\tau \hat\Lambda^0_0(\hat\xi)\hat X^i,\qquad [\hat X^i,\hat
X^j]=0,\label{newcoor}
\eeq
which, in turn,   can be interpreted as a generalization of
the $\kappa$-Minkowski space with a ``variable"
Planck length $\tau'=\tau \hat\Lambda^0_0(\hat\xi)$ that does
depend on {\em all} the quantum boost parameters (in the $(1+1)$D case,
this yields $\tau'=\tau \cosh\hat\xi$).
This result is a direct consequence  of imposing   a larger quantum group
symmetry than Poincar\'e.

In closing, I want to stress that Ref.~[\refcite{spainMink}] also
presents two maps that  can be used   to express
 the same quantum deformation of $so(4,2)$  within two physically different
frameworks: $U_\tau( {\mathcal CM}^{3+1})\leftrightarrow  U_{\tau}( AdS^{4+1})$.
In fact, such a quantum group relationship might  further be applied in order to
analyze the role that quantum deformations of $so(4,2)$ could play  in relation
with   the   ``AdS-CFT  correspondence" that relates local QFT on $AdS^{(d-1)+1}$
with a conformal QFT on
the  (compactified) Minkowskian spacetime ${\mathcal CM}^{(d-2)+1}$.


\begin{thebibliography}{0}

\bibitem{LukierskiRuegg1992} J.~Lukierski, A.~Nowicki,
H.~Ruegg,   V.N.~Tolstoy,
Phys. Lett. B 264 (1991) 331;
S.~Giller, P.~Kosinski,
J.~Kunz,  M.~Majewski,  P.~Maslanka,
Phys. Lett. B 286(1992)57;
J.~Lukierski,  H.~Ruegg,   A.~Nowicky,
Phys Lett. B {293}  (1992) 344.

\bibitem{nulla}
A.~Ballesteros,  F.J.~Herranz, M.A.~del Olmo, M.~Santander,
{Phys. Lett. B} {351}(1995)137;
A.~Ballesteros,  F.J.~Herranz, C.M.~Pere\~na,
{Phys. Lett. B} {391}  (1997) 71.

\bibitem{Majida}
 S.~Majid,   {Class. Quantum Grav.} {5} 1587 (1988).

  \bibitem{Amelino-Camelia:2000}
G.~Amelino-Camelia,
Phys.\ Lett.\ B { 510} (2001) 255;
 Int.\ J.\ Mod.\ Phys.\ D {11} (2002) 35.

\bibitem{MagueijoSmolin}
J.~Magueijo,  L.~Smolin,
Phys.~Rev.~Lett.~88 (2002) 190403.

\bibitem{Amelino-Camelia:2002vy}
G.~Amelino-Camelia,
Int.\ J.\ Mod.\ Phys.\ D {11} (2002) 1643;
Nature 418 (2002) 34.

 \bibitem{Kowalski-Glikman:2002we}
J.~Kowalski-Glikman,  S.~Nowak,
Phys.\ Lett.\ B {539} (2002) 126.

\bibitem{Bruno:primo}
N.R.~Bruno, G.~Amelino-Camelia,  J.~Kowalski-Glikman,   Phys.\ Lett.\ B {522} (2001) 133.

\bibitem{Bruno:2002wc}
N.R.~Bruno,
Phys.\ Lett.\ B {547} (2002) 109.

\bibitem{Lukierski:2002df}
J.~Lukierski, A.~Nowicki,
Int.\ J.\ Mod.\ Phys.\ A {18} (2003) 7.

\bibitem{spainDSR}
A.~Ballesteros, N.R.~Bruno, F.J.~Herranz,
 J. Phys. A 36 (2003) 10493.

\bibitem{spainMink}
A.~Ballesteros, N.R.~Bruno, F.J.~Herranz,
 Phys. Lett B 574 (2003) 276.

\bibitem{Herranz:2002fe}
F.J.~Herranz,
Phys.\ Lett.\ B {543} (2002) 89.

\bibitem{Maslanka} P.~Maslanka, J.\ Phys.\ A {26} (1993) L1251.

\bibitem{Majid:1994cy}
S.~Majid, H.~Ruegg,
Phys.\ Lett.\ B {334} (1994) 348.

\bibitem{Zak1} S.~Zakrzewski, J.\ Phys.\ A {27} (1994) 2075.

\bibitem{Zak2} J.~Lukierski, H.~Ruegg,
Phys.\ Lett.\ B {329} (1994) 189.

\bibitem{gacdual} A.~Agostini, G.~Amelino-Camelia,
F.~D'Andrea,
%HOPF ALGEBRA DESCRIPTION OF NONCOMMUTATIVE SPACE-TIME SYMMETRIES.
hep-th/0306013.

\bibitem{Lukierskid}
J.~Lukierski,V.D.~Lyakhovsky, M.~Mozrzymas,
{Phys. Lett. B}  {538} (2002) 375.

\bibitem{Faddeev:ih}
L.D.~Faddeev, N.Y.~Reshetikhin, L.A.~Takhtajan,
Lengingrad Math.\ J.\  {1} (1990) 193.

\end{thebibliography}
\end{document}